\documentclass[11pt]{article}

\usepackage[T1]{fontenc}
\usepackage[utf8]{inputenc}
\usepackage{lmodern}
\usepackage{textcomp}
\usepackage[margin=1in]{geometry}
\usepackage{graphicx}
\usepackage{booktabs}
\usepackage{longtable}
\usepackage{array}
\usepackage{calc}
\usepackage{authblk}
\usepackage{amsmath,amssymb}
\usepackage{microtype}
\usepackage[hidelinks,colorlinks=true,urlcolor=blue,linkcolor=blue,citecolor=blue]{hyperref}
\usepackage{url}
\usepackage{xcolor}
\usepackage{pifont}

\renewcommand{\checkmark}{\ding{51}}

\title{PyPeakRankR: Reproducible Peak-Level Feature Extraction for Regulatory Element Ranking}

\author[1,*]{Saroja Somasundaram}
\author[1]{Nelson J. Johansen}
\author[1]{Trygve E. Bakken}
\author[1,*]{Jeremy A. Miller}

\affil[1]{Allen Institute for Brain Science, Seattle, WA, USA}
\affil[*]{Corresponding authors}

\date{}

\begin{document}
\maketitle

\begin{abstract}
\noindent High-throughput chromatin accessibility assays such as ATAC-seq generate thousands of candidate regulatory elements (peaks), yet no standardized tool exists for assembling the diverse quantitative features needed to prioritize peaks for functional validation. Here we present PyPeakRankR, an open-source Python package that extracts peak-level features, namely BigWig signal summaries, GC content, PhyloP conservation scores, distribution moments (kurtosis, skewness, bimodality), and cell-type specificity rankings, into a single reproducible peak by feature matrix stored as a tab-separated values (TSV) file. PyPeakRankR separates deterministic feature extraction from downstream ranking, enabling transparent benchmarking of prioritization strategies on the same upstream data. The package provides both a command-line interface and a matching Python API, supports cross-assembly scoring via liftOver, and runs in minutes on thousands of peaks. PyPeakRankR was validated in the Brain Initiative Cell Census Network (BICCN) community challenge, where its predecessor PeakRankR ranked among the top 3 of 16 methods for cell-type-specific enhancer prediction. In a recent basal ganglia study, PyPeakRankR was used within the Cross-species Enhancer Ranking Pipeline (CERP) to identify enhancer-AAV tools achieving greater than 70\% on-target specificity across cell types. PyPeakRankR is freely available under the MIT license at \url{https://github.com/AllenInstitute/PeakRankR/tree/python-package}.

\medskip
\noindent\textbf{Keywords:} ATAC-seq, genomics, regulatory elements, bioinformatics, peak ranking, enhancer, Python
\end{abstract}

\section{Introduction}\label{introduction}

High-throughput chromatin accessibility assays such as ATAC-seq (\hyperref[ref-Buenrostro2013]{Buenrostro et al.~2013}) generate large sets of candidate regulatory elements called ``peaks.'' Downstream analyses require prioritizing peaks across cell types, experimental conditions, or species to identify the most biologically relevant candidates for functional validation. However, peak prioritization workflows are frequently implemented using ad hoc scripts with inconsistent feature definitions and aggregation strategies, limiting reproducibility and cross-study comparability.

Prioritizing genomic peaks across cell types requires combining multiple features: signal intensity, sequence properties such as GC content, evolutionary conservation (PhyloP (\hyperref[ref-Pollard2010]{Pollard et al.~2010})), and higher-order signal statistics. These features are typically computed using custom per-project scripts that vary across laboratories, complicating benchmarking and cross-study integration. This gap particularly affects computational biologists working with single-cell ATAC-seq (scATAC-seq) or bulk ATAC-seq data who need to systematically prioritize peaks for experimental follow-up, especially for enhancer discovery or adeno-associated virus (AAV) tool design.

Existing genomics tools each address part of the problem but none assembles a unified, portable feature matrix. Peak callers such as MACS2 (\hyperref[ref-Zhang2008]{Zhang et al.~2008}) identify open chromatin regions but rank peaks only by fold change or p-value, reflecting signal strength rather than cell-type specificity. A peak with high fold change may be active across many cell types and therefore a poor candidate for cell-type targeted AAV tools. Differential accessibility tools such as ArchR (\hyperref[ref-Granja2021]{Granja et al.~2021}) test for cell-type enrichment but operate within their own data model and do not produce portable, tool-agnostic feature tables. Annotation tools such as GREAT (\hyperref[ref-McLean2010]{McLean et al.~2010}) link peaks to genes but do not score chromatin features. At the library level, pyBigWig (\hyperref[ref-Ramirez2020pyBigWig]{Ram\'irez and Diehl 2020}) provides low-level BigWig access without peak-level aggregation, deepTools (\hyperref[ref-Ramirez2016deepTools]{Ram\'irez et al.~2016}) computes matrix summaries oriented toward visualization, and pyfaidx (\hyperref[ref-Shirley2015]{Shirley et al.~2015}) enables FASTA access without a genomics feature pipeline.

Table~\ref{tab:features} summarises feature coverage across these tools.

\begin{table}[h]
\centering
\small
\setlength{\tabcolsep}{4pt}
\begin{tabular}{@{}lcccccccc@{}}
\toprule
Tool & \shortstack{Peak\\signal} & \shortstack{Cell-type\\specificity} & \shortstack{GC\\content} & \shortstack{PhyloP\\cons.} & \shortstack{Signal\\moments} & \shortstack{Portable\\TSV} & \shortstack{CLI +\\Python API} & \shortstack{Cross-\\assembly} \\
\midrule
PyPeakRankR & \checkmark & \checkmark & \checkmark & \checkmark & \checkmark & \checkmark & \checkmark & \checkmark \\
ArchR       & \checkmark & \checkmark & --         & --         & --         & partial    & --         & -- \\
MACS2       & \checkmark & --         & --         & --         & --         & \checkmark & \checkmark & -- \\
deepTools   & \checkmark & partial    & --         & --         & --         & \checkmark & \checkmark & -- \\
GREAT       & --         & --         & --         & --         & --         & \checkmark & web only   & -- \\
pyBigWig    & low-level  & --         & --         & --         & --         & --         & \checkmark & -- \\
pyfaidx     & --         & --         & via FASTA  & --         & --         & --         & \checkmark & -- \\
\bottomrule
\end{tabular}
\caption{Feature coverage across genomics tools. \checkmark = supported natively; partial = limited or indirect support; -- = not supported. ArchR portable output is partial because outputs are tied to the ArchR project object.}
\label{tab:features}
\end{table}

PyPeakRankR fills this gap by combining pyBigWig, pyfaidx, and SciPy (\hyperref[ref-Virtanen2020]{Virtanen et al.~2020}) into a flexible CLI pipeline that assembles heterogeneous features into a single reproducible TSV table.

\section{Software design}\label{software-design}

PyPeakRankR is built around three core design decisions:

\paragraph{Table-first, flexible pipeline.} Every subcommand reads an existing tab-separated values (TSV) file and appends columns without modifying the peak coordinates. Users can run any subset of steps or add custom columns; the table remains valid throughout. The table-first design enables incremental extension.

\paragraph{Separation of feature extraction from ranking.} Feature extraction is deterministic: given the same inputs and the same BigWig files, the same table is always produced. Ranking is deliberately left to the user or to the \texttt{rank-specificity} subcommand, which implements one well-defined ranking formula but is not the only option. This separation means benchmarking studies can compare ranking strategies using the same upstream feature matrix, which is exactly how PyPeakRankR was used in the Brain Initiative Cell Census Network (BICCN) challenge (\hyperref[ref-Johansen2025]{Johansen et al.~2025}).

\paragraph{Command-line interface (CLI) + Python API parity.} Every subcommand wraps a public Python function (\texttt{init\_table}, \texttt{add\_signal}, \texttt{add\_gc}, \texttt{add\_phylop}, \texttt{add\_moments}, \texttt{rank\_by\_specificity}), so the tool works equally in shell pipelines and Python notebooks without reimplementing logic.

The specificity ranking formula computes the ratio of target group signal to mean background signal, then min-max normalises to $[0, 1]$. This matches the CERP pipeline (\hyperref[ref-Wirthlin2026]{Wirthlin et al.~2026}) and the ATAC-specificity metric validated in the BICCN challenge (\hyperref[ref-Johansen2025]{Johansen et al.~2025}).

Each feature has a distinct biological rationale. GC content is lower in active enhancers than in promoters or bulk genomic DNA, reflecting differences in nucleosome occupancy. PhyloP conservation (\hyperref[ref-Pollard2010]{Pollard et al.~2010}) identifies peaks under cross-species purifying selection. Signal distribution moments, including kurtosis (sharpness), skewness (asymmetry), and bimodality (Sarle's coefficient), are motivated by Lu et al.\ (\hyperref[ref-Lu2015]{Lu et al.~2015}), who showed these shape features distinguish enhancers from promoters in ChIP-seq data more reliably than signal intensity alone. The table-first design is directly extensible: future columns could include sequence model importance scores (e.g., from Borzoi or Enformer) or spatially resolved scores from multiplexed error-robust fluorescence in situ hybridization (MERFISH), integrating epigenomic and spatial context in one reproducible matrix.

\begin{figure}[h]
\centering
\includegraphics[width=0.85\linewidth]{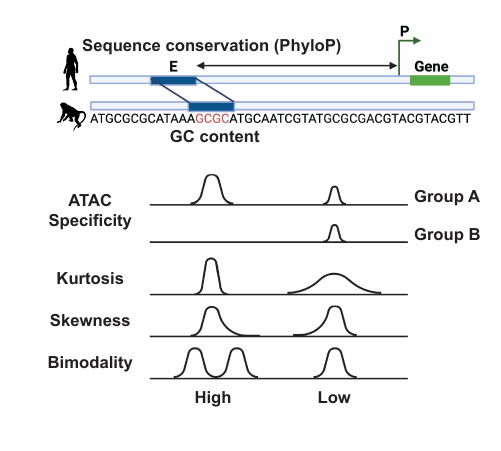}
\caption{Features collected by PyPeakRankR for each candidate peak: GC content, PhyloP conservation, ATAC specificity, and signal distribution moments (kurtosis, skewness, bimodality). Figure adapted from Wirthlin et al.~(\hyperref[ref-Wirthlin2026]{2026}).}
\label{fig:features}
\end{figure}

\section{Implementation}\label{implementation}

PyPeakRankR is implemented in Python ($\geq$3.9) with the following dependencies: pandas (\hyperref[ref-Reback2020]{pandas development team 2020}) for tabular data handling, NumPy (\hyperref[ref-Harris2020]{Harris et al.~2020}) for numerical computation, pyBigWig (\hyperref[ref-Ramirez2020pyBigWig]{Ram\'irez and Diehl 2020}) for BigWig signal extraction, pyfaidx (\hyperref[ref-Shirley2015]{Shirley et al.~2015}) for FASTA sequence access, and SciPy (\hyperref[ref-Virtanen2020]{Virtanen et al.~2020}) for statistical distribution metrics. The package is installable via pip from GitHub and includes a \texttt{pypeakranker} CLI, unit tests, and example data. Source code is available at \url{https://github.com/AllenInstitute/PeakRankR/tree/python-package} under the MIT license.

\section{Results}\label{results}

\subsection{BICCN community challenge validation}\label{biccn-community-challenge-validation}

PyPeakRankR extends the R package PeakRankR, which used a minimal set of three features (ATAC specificity, signal magnitude, and peak coverage) in the BICCN Community Challenge (\hyperref[ref-Johansen2025]{Johansen et al.~2025}). PeakRankR ranked among the top 3 of 16 methods for predicting cell-type-specific enhancers in the mammalian cortex. PyPeakRankR re-implements and expands this approach in Python to integrate directly with sequence models and modern genomics workflows.

\subsection{Cross-species enhancer-AAV toolkit}\label{cross-species-enhancer-aav-toolkit}

In a recent basal ganglia study from our group (\hyperref[ref-Wirthlin2026]{Wirthlin et al.~2026}), PyPeakRankR was used within the Cross-species Enhancer Ranking Pipeline (CERP) across multiple basal ganglia (BG) cell types in mouse and macaque. The composite feature rankings outperformed conventional fold-change approaches, and the resulting enhancer-AAV tools achieved $>$70\% on-target specificity across cell types, with exemplary enhancers exceeding 90\%.

These results establish direct experimental utility of the feature extraction and ranking approach implemented in PyPeakRankR.

\section{Discussion}\label{discussion}

PyPeakRankR addresses the lack of a standardized, portable tool for peak-level feature extraction in regulatory genomics. By separating deterministic feature computation from ranking logic, PyPeakRankR enables reproducible benchmarking of prioritization strategies. Its validation in the BICCN community challenge and application in cross-species enhancer-AAV design demonstrate practical utility for enhancer discovery workflows.

The table-first design is directly extensible: future columns could include sequence model importance scores (e.g., from Borzoi or Enformer) or spatially resolved scores from MERFISH, integrating epigenomic and spatial context in one reproducible matrix.

\section{Data and code availability}\label{data-and-code-availability}

PyPeakRankR is open source (MIT license) and available at \url{https://github.com/AllenInstitute/PeakRankR/tree/python-package}. The Zenodo archive is available at \url{https://doi.org/10.5281/zenodo.15238527}.

\section*{Acknowledgements}

The authors thank the bioinformatics and enhancer adeno-associated virus (AAV) teams at the Allen Institute for Brain Science for feedback on feature definitions and pipeline design.

This research was supported by the Allen Institute, founded by Jody Allen, chair and co-founder of Allen Family Philanthropies, and the late Paul G. Allen, investor, philanthropist, and co-founder of Microsoft. We gratefully acknowledge their vision and generosity, which make this work possible. This research was also supported by U.S. National Institutes of Health (NIH) BRAIN Initiative Human and Mammalian Brain Atlas (HMBA) BICAN grant UM1MH130981. The content of this study is solely the responsibility of the authors and does not necessarily represent the official views of the National Institutes of Health.

\section*{Author contributions}

\textbf{Saroja Somasundaram:} Software development, methodology, writing, original draft. \textbf{Nelson J. Johansen:} Conceptualization, validation (BICCN challenge). \textbf{Trygve E. Bakken:} Supervision, writing, review and editing. \textbf{Jeremy A. Miller:} Supervision, conceptualization, writing, review and editing.

\section*{Competing interests}

The authors declare no competing interests.

\section*{References}
\begin{description}\setlength{\itemsep}{2pt}

\item[\phantomsection\label{ref-Buenrostro2013}] Buenrostro, J.~D., Giresi, P.~G., Zaba, L.~C., Chang, H.~Y., \& Greenleaf, W.~J.\ (2013). Transposition of native chromatin for fast and sensitive epigenomic profiling of open chromatin, DNA-binding proteins and nucleosome position. \emph{Nature Methods}, 10(12), 1213--1218. \url{https://doi.org/10.1038/nmeth.2688}

\item[\phantomsection\label{ref-Granja2021}] Granja, J.~M., Corces, M.~R., Pierce, S.~E., et al.\ (2021). ArchR is a scalable software package for integrative single-cell chromatin accessibility analysis. \emph{Nature Genetics}, 53(3), 403--411. \url{https://doi.org/10.1038/s41588-021-00790-6}

\item[\phantomsection\label{ref-Harris2020}] Harris, C.~R., Millman, K.~J., van der Walt, S.~J., et al.\ (2020). Array programming with NumPy. \emph{Nature}, 585(7825), 357--367. \url{https://doi.org/10.1038/s41586-020-2649-2}

\item[\phantomsection\label{ref-Johansen2025}] Johansen, N.~J., Kempynck, N., Zemke, N.~R., Somasundaram, S., De Winter, S., et al.\ (2025). Evaluating methods for the prediction of cell-type-specific enhancers in the mammalian cortex. \emph{Cell Genomics}, 5(6), 100879. \url{https://doi.org/10.1016/j.xgen.2025.100879}

\item[\phantomsection\label{ref-Lu2015}] Lu, Y., Qu, W., Shan, G., \& Zhang, C.\ (2015). DELTA: A distal enhancer locating tool based on AdaBoost algorithm and shape features of chromatin modifications. \emph{PLoS ONE}, 10(6), e0130622. \url{https://doi.org/10.1371/journal.pone.0130622}

\item[\phantomsection\label{ref-McLean2010}] McLean, C.~Y., Bristor, D., Hiller, M., et al.\ (2010). GREAT improves functional interpretation of cis-regulatory regions. \emph{Nature Biotechnology}, 28(5), 495--501. \url{https://doi.org/10.1038/nbt.1630}

\item[\phantomsection\label{ref-Pollard2010}] Pollard, K.~S., Hubisz, M.~J., Rosenbloom, K.~R., \& Siepel, A.\ (2010). Detection of nonneutral substitution rates on mammalian phylogenies. \emph{Genome Research}, 20(1), 110--121. \url{https://doi.org/10.1101/gr.097857.109}

\item[\phantomsection\label{ref-Ramirez2020pyBigWig}] Ram\'irez, F., \& Diehl, S.\ (2020). \emph{pyBigWig: A Python extension for reading BigWig files}. \url{https://github.com/deeptools/pyBigWig}

\item[\phantomsection\label{ref-Ramirez2016deepTools}] Ram\'irez, F., Ryan, D.~P., Gr\"uning, B., et al.\ (2016). deepTools2: A next generation web server for deep-sequencing data analysis. \emph{Nucleic Acids Research}, 44(W1), W160--W165. \url{https://doi.org/10.1093/nar/gkw257}

\item[\phantomsection\label{ref-Shirley2015}] Shirley, M.~D., Ma, Z., Pedersen, B.~S., \& Wheelan, S.~J.\ (2015). \emph{Efficient ``Pythonic'' access to FASTA files using pyfaidx}. \url{https://doi.org/10.7287/peerj.preprints.970v1}

\item[\phantomsection\label{ref-Reback2020}] pandas development team.\ (2020). pandas-dev/pandas: Pandas. \url{https://doi.org/10.5281/zenodo.3509134}

\item[\phantomsection\label{ref-Virtanen2020}] Virtanen, P., Gommers, R., Oliphant, T.~E., et al.\ (2020). SciPy 1.0: Fundamental algorithms for scientific computing in Python. \emph{Nature Methods}, 17(3), 261--272. \url{https://doi.org/10.1038/s41592-019-0686-2}

\item[\phantomsection\label{ref-Wirthlin2026}] Wirthlin, M.~E., Hunker, A.~C., Somasundaram, S., et al.\ (2026). A cross-species enhancer-AAV toolkit for cell type-specific targeting across the basal ganglia. \emph{bioRxiv}, ahead of print. \url{https://doi.org/10.64898/2026.02.23.706695}

\item[\phantomsection\label{ref-Zhang2008}] Zhang, Y., Liu, T., Meyer, C.~A., et al.\ (2008). Model-based analysis of ChIP-seq (MACS). \emph{Genome Biology}, 9(9), R137. \url{https://doi.org/10.1186/gb-2008-9-9-r137}

\end{description}

\end{document}